\documentstyle[12pt,epsfig]{article}
\def\thebibliography#1{\leftline{\large \it References}\list
  {[\arabic{enumi}]}{\settowidth\labelwidth{[#1]}\leftmargin\labelwidth
    \advance\leftmargin\labelsep
    \usecounter{enumi}}
    \def\newblock{\hskip .11em plus .33em minus .07em}
    \sloppy\clubpenalty4000\widowpenalty4000}

\newcommand{\be}{\begin{eqnarray}}
\newcommand{\ba}{\begin{array}}
\newcommand{\ea}{\end{array}}
\newcommand{\ee}{\end{eqnarray}}

\begin{document}
\rightline{UNITU-THEP-7/1996}
\rightline{SU-4240-631}
\rightline{May 1996}
%\rightline{hep-ph/9604nnn}
\vskip 1.5truecm
\centerline{\Large\bf Scaling Behavior in Soliton Models}
\vskip 0.5truecm
\baselineskip=18 true pt
\vskip 1.0cm
\centerline{M.\ Harada$^{a)}$, F.\ Sannino$^{a,b)}$, 
J.\ Schechter$^{a)}$, and H.\ Weigel$^{c)}$}
\vskip 0.5cm
\centerline{$^{a)}$Physics Department, Syracuse University}
\centerline{Syracuse, NY 13244--1130, USA}
\vskip 0.3cm
\centerline{$^{b)}$Dipartimento di Scienze Fisiche \& Istituto 
Nazionale di Fisica Nucleare}
\centerline{Mostra D'Oltremare Pad. 19,  80125 Napoli, Italy}
\vskip 0.3cm
\centerline{$^{c)}$Institute for Theoretical Physics, 
T\"ubingen University}
\centerline{Auf der Morgenstelle 14, D-72076 T\"ubingen, Germany}
\baselineskip=18pt
\vskip 3cm
\centerline{\bf ABSTRACT}
\vskip1cm
In the framework of chiral soliton models we study the behavior 
of static nucleon properties under rescaling of the parameters
describing the effective meson theory. In particular we investigate 
the question of whether the Brown--Rho scaling laws are general 
features of such models. When going beyond 
the simple Skyrme model we find that restrictive constraints 
need to be imposed on the mesonic parameters in order to maintain 
these scaling laws. Furthermore, in the case when vector mesons are
included in the model it turns out that the isoscalar form factor no
longer scales according to these laws. Finally we note that, in
addition to the exact scaling laws of the model, one may construct
approximate {\it local scaling laws}, which depend of the particular
choice of Lagrangian parameters.

\vfil\eject
\baselineskip=14pt
\leftline{\large\it 1. Introduction}
\medskip
The properties of nucleons in nuclear matter (at {\it finite density}
) are of great interest. Recently, these properties have been
intensively studied \cite{Br91,Br95} in the framework of a simple
soliton model to the baryon. Such models (supported by the $1/N_C$
expansion in QCD) describe nucleons in terms of effective chiral
Lagrangians of mesons. The central idea of these investigations is the
assumption that the properties of the nucleon at finite density are to
be obtained just by using different values of the parameters in the
underlying meson Lagrangian. The relationship between these parameters
and their zero density values can, for example, be estimated by using
QCD sum rules \cite{Ji95}--\cite{Co95}
Here we will not question the central assumption but rather will study
 whether the 
scaling laws, which hold in the simple Skyrme model, remain 
valid when more complicated but also more realistic soliton models are 
considered.

The typical results of this approach are the Brown-Rho scaling laws
\cite{Br91,Br95}, which hold exactly in the semi-classical treatment
of the Skyrme model with no pion mass term. These express the
invariance of the combinations:
\be
A & = & \frac{M}{f_{\pi} \sqrt{g_{A}}} =  \frac{M^{\ast}}{f^*_{\pi}
\sqrt{g^*_{A}}} \nonumber \ , \\
B  &=&  \frac{{\langle r^2 \rangle}_{I=0} {f_{\pi}}^2}{g_A} =
\frac{\langle r^2\rangle_{I=0}^{\ast} {f^{\ast}_{\pi}}^2}{g^{\ast}_A}
\ ,
\label{defAB1}
\ee
where the starred quantities correspond to evaluations at finite
density. $M$, $g_A$ and $\langle r^2 \rangle _{I=0}$ correspond to the
nucleons' mass, axial coupling constant and isoscalar squared
radius. The quantity $f_{\pi}$, which is present in the meson
Lagrangian, is the pion decay constant. 

It is natural to ask whether these equations hold in the simplest
extension of the Skyrme model in which a pion mass term is added. We
will show they do hold, {\it provided} that, in addition, $m_{\pi}$
scales like
\be
m_{\pi}^2 \langle r^2 \rangle _{I=0} = 
{m^{\ast}_{\pi}}^2 \langle r^2 \rangle _{I=0}^{\ast} \ .
\ee

Now, it is well known that the simple Skyrme model is unable to provide
an adequate description of several nucleon properties. Examples are: the 
non--electromagnetic piece of the neutron proton mass difference 
\cite{Ja89}, the proton matrix element of the axial singlet current 
\cite{Jo90} and the {\it high energy} behavior of the phase shifts in 
pion nucleon scattering \cite{Ec86}. All these short--comings can be 
improved in soliton models which contain explicit vector meson 
degrees of freedom. It is therefore natural to study the scaling 
behavior of static baryon properties in  a more realistic vector meson 
model\footnote{The density dependence of static properties was 
previously studied in such a model \cite{Me89} where 
the variation of the meson parameters was adopted from 
Nambu--Jona--Lasinio model calculations.}.

We will show that in this model the quantity $B$ in (\ref{defAB1})
is no longer invariant under rescaling. However the combination $A$ in
(\ref{defAB1}) will be invariant provided that 
\be
\frac{m_{\pi}}{m_{\rho}}=\frac{m_{\pi}^{\ast}}{m_{\rho}^{\ast}} 
\quad {\rm and}\quad 
\frac{m_{\rho}}{g f_{\pi}}=\frac{m_{\rho}^{\ast}}{g^{\ast} f_{\pi}^{\ast}} \ ,
\label{def2}
\ee
hold, as well as some scaling laws for coupling constants in the terms
of the meson Lagrangian proportional to the Levi--Civita symbol. In
(\ref{def2}) $g$ is the vector meson coupling constant and
$m_{\rho}$ is the $\rho$--meson mass.

The general strategy for obtaining these scaling laws is to first 
construct a universal expression for the soliton mass by absorbing
as many parameters of the effective meson theory as possible into 
 field and coordinate redefinitions. In this process not all meson
fields can be reparametrized because they are subject to restrictive 
boundary conditions. Subsequently these redefinitions are employed 
to study the behavior of other nucleon properties when varying these 
parameters.

The scaling laws mentioned above are exact consequences of the models
for any choice of parameters. It is also possible to find approximate
(local) scaling laws which hold in the vicinity of a given
parameter. An illustration is provided for the vector meson model. 

Tests of various scaling laws should be available after the RHIC
facility is completed.

\bigskip
\leftline{\large\it 2. The Pseudoscalar Model}
\medskip

Although the scaling laws for dense matter were motivated 
\cite{Br91} from an effective meson Lagrangian \cite{Go86}, which 
includes the QCD trace anomaly, these laws essentially follow from the 
simple Skyrme model \cite{Sk61,Ad83}. In order to present this model 
it is convenient to consider the root, $\xi$ of the non--linear 
realization, 
$\xi^2=U= {\rm exp}(i\mbox{\boldmath $\tau$}\cdot
\mbox{\boldmath $\pi$}/f_\pi)$,
of the pseudoscalar fields. Using the vector and pseudovector 
objects constructed from $\xi$ 
\be
v_\mu=\partial_\mu\xi\xi^{\dag}-\xi^{\dag}\partial_\mu\xi
\qquad {\rm and} \qquad
p_\mu=\partial_\mu\xi\xi^{\dag}+\xi^{\dag}\partial_\mu\xi\ . 
\label{defvp}
\ee
the Skyrme Lagrangian assumes the simple form
\be
{\cal L}_{\rm Sk}=-\frac{f_\pi^2}{4}{\rm tr}\left(p_\mu p^\mu\right)
+\frac{1}{32e^2}{\rm tr}\left(\left[p_\mu,p_\nu\right]
\left[p^\mu,p^\nu\right]\right)\ .
\label{lsk}
\ee
Substituting the hedgehog configuration 
$\xi(\mbox{\boldmath $r$})={\rm exp}(i\mbox{\boldmath $\tau$}\cdot
{\hat{\mbox{\boldmath $r$}}}F(r)/2)$ yields the classical energy 
functional
\be
E_{\rm cl}[F]&=&2\pi\frac{f_\pi}{e}\int_0^\infty \ d\zeta\Bigg\{
\zeta^2F^{\prime2}+2{\rm sin}^2F
+{\rm sin}^2F\left(2F^{\prime2}
+\frac{{\rm sin}^2F}{\zeta^2}\right) \Bigg\} \ ,
\label{ecl}
\ee
where the dimensionless variable $\zeta=ef_\pi r$ has been introduced.
This functional is the building block for discussing the variation of 
static nucleon properties with the parameters $f_\pi$ and $e$. 
The chiral angle, $F(\zeta)$ is obtained by minimizing $E_{\rm cl}[F]$ 
subject to the boundary conditions $F(0)=-\pi$ and $F(\infty)=0$ for 
topological baryon 
number one. Within this treatment one derives scaling formulas for 
the classical mass, the isoscalar radius, $\langle r^2\rangle_{I=0}$, 
and the axial vector charge, $g_A$ \cite{Ad83}
\be
E_{\rm cl}=72.9\frac{f_\pi}{e}\ , \quad
\langle r^2\rangle_{I=0}=\frac{1.12}{f_\pi^2 e^2} 
\quad {\rm and}\quad g_A=\frac{18.0}{e^2}\ .
\label{funiv}
\ee
It should be noted that in the simple Skyrme model the isoscalar 
radius receives its sole contribution from the topological baryon 
density. This current can be interpreted as the $U_V(1)$ Noether 
current from the Wess--Zumino term \cite{Wi83}
\be
\Gamma_{\rm WZ}=\frac{iN_C}{240\pi^2}\int_{M_5}
{\rm tr}(p^5)\ , \quad p=p_{\mu} dx^{\mu} \ , 
\label{gwz}
\ee
where the differential forms notation has been adopted.
Furthermore $M_5$ refers to a manifold which has Minkowski space, 
$M_4$, as boundary. We are assuming the three flavor case in writing 
(\ref{gwz}).

In leading order of the $1/N_C$ expansion the nucleon mass, $M$, is
identical to $E_{\rm cl}$. Hence by eliminating the Skyrme parameter, 
$e$, from the expressions (\ref{funiv}) one easily obtains 
two quantities which only contain physical observables 
\be
A:=\frac{M}{f_\pi\sqrt{g_A}}=17.2
\quad {\rm and}\quad
B:=\frac{\langle r^2\rangle_{I=0}f_\pi^2}{g_A}=0.062 \ ,
\label{defAB}
\ee
where the data refer to the Skyrme model predictions (\ref{funiv}).
In particular, these combinations are universal in the Skyrme model, 
in the sense that $A$ and $B$ remain unchanged when scaling the 
fundamental parameters of the Lagrangian (\ref{lsk}) 
\cite{Br91,Rho85}
\be
A^{\ast}=A \quad {\rm and}\quad B^{\ast}=B \ .
\label{brscal}
\ee
Here the asterisk denotes quantities, which are computed 
using\footnote{Although we do not specifically have in mind to
restrict the discussion to the 
density behavior of the nucleon properties we follow the 
conventions of dense matter literature and denote quantities which 
are associated with modified meson parameters by an asterisk.}
$f_\pi^{\ast}\ne f_\pi$ and $e^{\ast}\ne e$.

It is an easy exercise to verify that $A$ and $B$ 
are universal quantities independent of which higher order (in 
derivatives) stabilizing term is adopted. The only requirement 
is the restriction to a ${\underline {\rm single}}$ stabilizing 
term.

As the Skyrme model represents the simplest version of an 
effective meson theory which supports soliton solutions, it is 
certainly incomplete. Hence the natural question arises whether 
the scaling laws (\ref{brscal}) hold in more realistic models.
The first extension, which comes into mind in this context, is 
the inclusion of a pion mass term. This neither changes the 
analytical form of $\langle r^2\rangle_{I=0}$ nor of $g_A$; however
it contributes to the soliton mass: 
\be
(4\pi f_\pi/e)(m_\pi/e f_\pi)^2\int_0^\infty 
d\zeta \zeta^2(1-{\rm cos}F) \ .
\ee
Hence a universal expression for the nucleon mass can only be 
obtained by demanding that $m_\pi/e f_\pi$ does not scale. This 
implies
\be
m_\pi^{\ast 2}\langle r^2\rangle_{I=0}^{\ast}\equiv
m_\pi^2\langle r^2\rangle_{I=0}\ .
\label{mpisc}
\ee
We thus see that the scaling laws (\ref{brscal}) do not simply follow 
from a chirally symmetric theory; rather their implementation yields 
relations among the scaling behaviors of various hadron observables. 
Note that this relation goes beyond dimensional analysis since the 
model contains two dimensional parameters: $f_\pi$ and $m_\pi$. 
From Nambu--Jona--Lasinio model studies on the density dependence of 
$m_\pi$ \cite{Be87} one might conclude that 
$\langle r^2\rangle_{I=0}^{\ast}$ does not vary until three times 
nucleon matter density is reached. In any event, the role of the pion 
mass is special as it arises from the explicit breaking of chiral 
symmetry. In the following sections we will therefore discuss such 
constraints for coefficients of chirally symmetric expressions in 
more realistic models which include vector mesons.

\bigskip
\leftline{\large\it 3. The Vector Meson Model}
\medskip

The chirally symmetric action for a realistic vector meson model 
has been worked out previously \cite{Ka84,Ja88}
\be
{\cal A}&=&\int d^4 x {\cal L}_{\rm nan}+
\Gamma_{\rm an}+\Gamma_{\rm WZ}
\label{vmact} \\
{\cal L}_{\rm nan}&=&{\rm tr}\left[
-\frac{f_\pi^2}{4}p_\mu p^\mu 
+\frac{m_\pi^2 f_\pi^2}{4}\left(U+U^{\dag}-2\right)
-\frac{1}{2}F_{\mu\nu}(\rho)F^{\mu\nu}(\rho)
+m_\rho^2 R_\mu R^\mu\right]
\nonumber \\
\Gamma_{\rm an}&=&\int{\rm tr}\left(
\frac{1}{6}\left[\gamma_1+\frac{3}{2}\gamma_2\right]Rp^3
-\frac{i}{4}g\gamma_2 F(\rho)\left[pR-Rp\right]
-g^2\left[\gamma_2+2\gamma_3\right]R^3p\right)\ .
\nonumber
\ee
For convenience the differential forms notation has again been used
to simplify the presentation of the $\epsilon_{\mu\nu\rho\sigma}$
terms. The vector mesons $\omega$ and $\rho$ are contained in 
$R_\mu=\rho_\mu-(i/2g)v_\mu$ with 
$\rho_\mu=(\omega_\mu+\rho_\mu^a\tau_a)/2$. 
The additional parameters $g$ and $\gamma_i$ are related to the decay 
widths of the vector mesons. Using a slightly different notation
they are found to be \cite{Ja88}
\be
g\approx 5.6\ , \quad 
{\tilde h}=-\frac{2\sqrt{2}}{3}\ \gamma_1\approx 0.4\ , \quad
{\tilde g}_{VV\phi}=g\gamma_2\approx 1.9 \ , \quad
\kappa=\frac{\gamma_3}{\gamma_2}\approx 1\ .
\label{vmpara}
\ee
The value for $\kappa$ was obtained from a fit to static nucleon 
properties.

In addition to the hedgehog {\it ansatz} for the pseudoscalar fields 
we have two more radial functions parametrizing the vector meson
fields
\be
\omega_0={{\omega(r)}\over{g} m_{V}}\ ,\quad
\rho_i^a={{G(r)}\over{gr}}\epsilon_{ija}\hat r_j\ ,
\label{vmhedgehog}
\ee
where $m_{V}=\sqrt{2} g f_{\pi}$.
Introducing a scaled radial coordinate $\zeta =m_{V}r$ 
yields the classical energy functional \cite{Ja88}
\be
E_{\rm cl}&=&\frac{4\pi f_\pi^2}{m_V}\int_0^\infty d\zeta
\Bigg\{\frac{1}{2}\left(F^{\prime2}\zeta^2+2{\rm sin}^2F\right)
+\frac{1}{4}\mu_\pi^2\zeta^2\left(1-{\rm cos}F\right)
-\left(\omega^{\prime2}+\mu_\rho^2\omega^2\right)\zeta^2
\nonumber \\ && \hspace{0.5cm}
+G^{\prime2}+\frac{G^2}{2\zeta^2}\left(G+2\right)^2
+\mu_\rho^2\left(1+G-{\rm cos}F\right)^2
+2g\gamma_1 F^\prime\omega{\rm sin}^2F
-4g\gamma_2 G^\prime\omega{\rm sin}F
\nonumber \\ && \hspace{1.0cm}
+2g\gamma_3 F^\prime\omega G\left(G+2\right)
+2g\left(\gamma_2+\gamma_3\right) F^\prime\omega
\left(1-2(G+1){\rm cos}F+{\rm cos}^2F\right)\Bigg\} .
\label{eclvm}
\ee
Here a prime denotes the derivative with respect to $\zeta$ and 
$\mu_i=m_i/m_V$ are dimensionless mass parameters. In order to obtain 
a universal mass functional we demand that none of the coefficients 
in eq (\ref{eclvm}) vary when $f_\pi$ and/or $g$ are scaled. 
This implies
\be
\frac{m_\pi^{\ast}}{g^{\ast}f_\pi^{\ast}}=
\frac{m_\pi}{g f_\pi}\ , \quad
\frac{m_\rho^{\ast}}{g^{\ast}f_\pi^{\ast}}=
\frac{m_\rho}{g f_\pi}\ ,\quad
g^{\ast}{\tilde h}^{\ast}=g{\tilde h}\ , \quad
{\tilde g}_{VV\phi}^{\ast}={\tilde g}_{VV\phi}\
\quad {\rm and} \quad \kappa^{\ast}=\kappa\ .
\label{vmscal}
\ee
The second equation states that for a universal mass functional 
the KSRF \cite{Ka66} relation has to be taken invariant. For 
$g^{\ast}\ne g$ the frequently adopted scaling relation 
\cite{Br91,Br96,Fr96,Rho96}
$m_\rho^{\ast}/m_\rho=f_\pi^{\ast}/f_\pi$ is apparently violated.

Static properties are computed as the appropriate matrix elements 
of the symmetry currents. In order to obtain these currents one 
first introduces external left and right gauge fields ($B_L^\mu$ 
and $B_R^\mu$) such that the action (\ref{vmact}) is invariant under 
local chiral transformations (up to the anomaly). This introduces 
an additional term, which does not contribute to the classical 
energy functional, in the anomalous sector \cite{Ja88}
\be
d_1\int {\rm tr}\left(\xi F(B_R) \xi^{\dag}+
\xi^{\dag}F(B_L)\xi\right)\left(Rp-pR\right) \ .
\label{d1term}
\ee
This part of the gauged action together with the $\gamma_2$--term 
contributes to the decay width of the process $\omega\to\pi^0\gamma$.
This determines 
$|2d_1-{\tilde g}_{VV\phi}/2g^2|\approx0.038$ \cite{Me89b}.
The currents are extracted from the expression linear in 
these gauge fields. Substituting the static field configurations into 
the so--obtained currents and taking matrix elements
provides the static nucleon properties. For the axial charge one 
finds \cite{Me89b}
\be
g_A&=&\frac{4\pi}{9g^2}\int_0^\infty d\zeta
\left(2\zeta a_1 + a_2\right)
\label{gavm} \\
a_1&=&{\rm sin}F{\rm cos}F
+2\mu_\rho^2{\rm sin}F\left(1+G+{\rm cos}F\right)
+g\left(2\gamma_1+3\gamma_2\right)
\omega F^\prime{\rm sin}F{\rm cos}F
\nonumber \\ &&
-g\gamma_2\left[{\rm cos}F
\left(\omega G^\prime-\omega^\prime(1+G-{\rm cos}F)\right)
-\omega^\prime{\rm sin}^2F\right]
\nonumber \\ &&
+g\left(\gamma_2+2\gamma_3\right)F^\prime\omega{\rm sin}F
(1+G-{\rm cos}F)
+4g^2d_1\left(\omega^\prime{\rm sin}^2F
+\omega F^\prime{\rm sin}2F\right)
\nonumber \\
a_2&=&\zeta^2F^\prime+g\left(2\gamma_1+3\gamma_2\right)
\omega {\rm sin}^2F -g\gamma_2\omega G(G+2)
\nonumber \\ &&
+g\left(\gamma_2+2\gamma_3\right)\omega
(1+G-{\rm cos}F)^2+8g^2d_1\omega {\rm sin}^2F\ .
\nonumber
\ee
Although the term proportional to $d_1$ is a total derivative 
and hence does not contribute to $g_A$, it is clear that 
demanding $g^{\ast2}d^{\ast}_1=g^2d_1$ in addition to the scaling 
laws (\ref{vmscal}) enables the higher moments of the axial current to
scale universally. Note that the first Brown--Rho scaling condition in
(\ref{brscal}) $A^{\ast}=A$ is satisfied.
Starting from the expression for the pion--nucleon 
coupling constant \cite{Me89b}
\be
g_{\pi NN}=\frac{8\pi}{9}Mf_\pi m_\pi^2
\int_0^\infty dr\ r^3 {\rm sin}F(r)
\label{gpinn}
\ee
it is straightforward to verify that 
$g^{\ast}_{\pi NN}f^{\ast}_\pi/M^{\ast}g^{\ast}_A=
g_{\pi NN}f_\pi/Mg_A$ when (\ref{vmscal}) is imposed, {\it i.e.}
the Goldberger Treiman relation is scale independent. From (\ref{gavm}) it can be seen that the Wess--Zumino 
term (\ref{gwz}) does not contribute to $g_A$. This is in contrast 
to the isoscalar radius \cite{Me89b}
\be
\langle r^2\rangle_{I=0}&=&\frac{8\pi}{m_V^2N_C}
\int_0^\infty d\zeta \zeta^2 R_{I=0}
\label{r2vm} \\
R_{I=0}&=&\frac{\mu_\rho}{g^2}\zeta^2\omega
+\left[\frac{N_C}{4\pi^2}-\frac{1}{g}
\left(\gamma_1+\frac{3}{2}\gamma_2\right)\right]F^\prime{\rm sin}^2F
+\frac{\gamma_2}{2g}\left[F^\prime G(G+2)-2G^\prime{\rm sin}F\right]
\nonumber \\ &&
-\frac{\gamma_2+2\gamma_3}{2g}F^\prime
(1+G-{\rm cos}F)^2
+4d_1\frac{d}{d\zeta}\left[{\rm sin}F(1+G-{\rm cos}F)\right] \ ,
\nonumber 
\ee
where the term which involves $N_C$ stems from $\Gamma_{\rm WZ}$.
This term scales differently from all the others and causes 
$\langle r^2\rangle_{I=0}$ 
${\underline{\rm not}}$ to scale universally. If the Wess--Zumino 
term had yielded the sole contribution the isoscalar radius would 
scale like $1/m^{\ast}_V\sim 1/g^{\ast}f_\pi^{\ast}$. As a consequence 
the quantity $B$, which is defined in eq (\ref{defAB}), would be scale 
independent and one would recover the simple Skyrme model result. 
Also the relation (\ref{mpisc}) would remain valid.
On the other hand, if the other terms, which represent the vector
contributions, were the only ones present
 the scale independent quantity would 
instead be
\be
{\tilde B}:=\frac{\langle r^2\rangle_{I=0}f_\pi^2}{g_A^2} \ .
\label{deftB}
\ee
In figure 1 we display the numerical comparison of the scale 
dependences of $B$ and ${\tilde B}$. For this study the variations of 
the meson parameters are described by the {\it ans\"atze}
\be
f_\pi^{\ast}=f_\pi(1-x)
\quad {\rm and} \quad
g^{\ast}=g(1-cx)\ .
\label{fgscal}
\ee
According to eq (\ref{vmscal}) this implies
$m_\rho^{\ast}\approx m_\rho(1-(1+c)x)$ for small $x$. With $c=-0.34$ 
we obtain 
$m_\rho^{\ast}/m_\rho\approx0.78$ and $g_A^{\ast}/g_A\approx0.80$ 
at $x=x_{\rm E}=0.35$ as suggested for nuclear density by QCD sum 
rules \cite{Ji95,Fu95} and studies of the neutron beta--decay in heavy 
nuclei \cite{La95}, respectively. Note that $c\ne0$ is necessary for 
$g_A$ to vary. 
\begin{figure}[ht]
\centerline{
\epsfig{figure=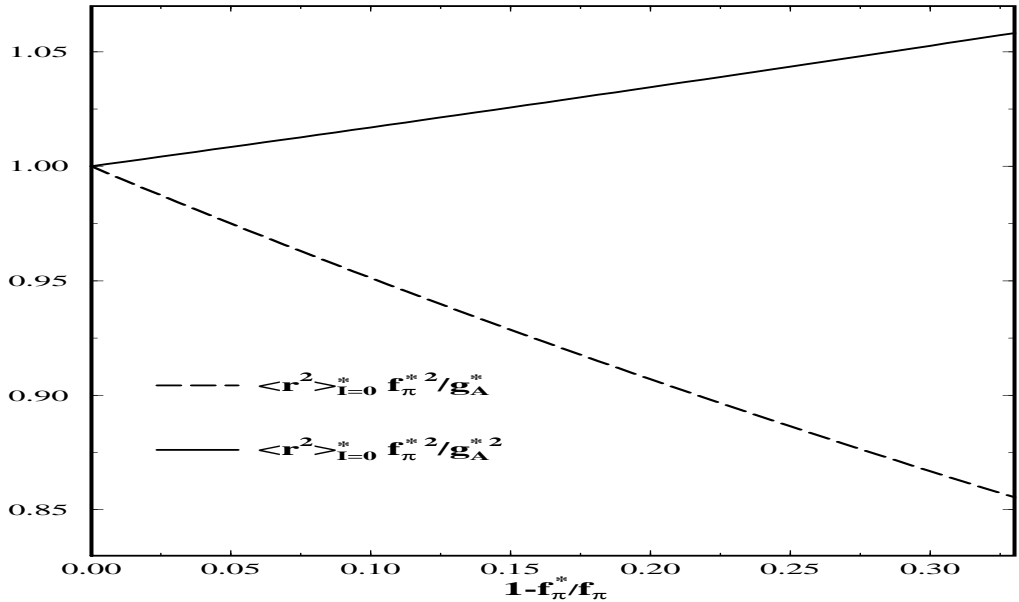,height=8.0cm,width=15.0cm}}
\vskip0.8cm
\caption{Comparison of the two scaling behaviors for the isoscalar 
radius in the vector meson model. These quantities are normalized 
to their unscaled values. The variation of the soliton properties
corresponds to $c=-0.34$ in eq (\protect\ref{fgscal}).}
\end{figure}
The numerical calculation yields $M^{\ast}/M\approx0.60$ at 
$x=x_{\rm E}$, which is slightly smaller than the QCD sum 
rules estimate of $0.67\pm0.05$ \cite{Ji95,Fu95}.

The scaling behavior shown in figure 1 indicates that ${\tilde B}$
 is closer to an invariant than $B$; the variation of the former is 
only about 5\% compared to a 17\% decrease of $B$ at $x_E$. The 
physical interpretation of this result is that the vector mesons
 provide the major 
contribution to the isoscalar form factor. This in turn alters the 
scaling law obtained in the simple Skyrme model which does not contain 
any vector meson degrees of freedom. On the other hand the isovector 
radius $\langle r^2\rangle_{I=1}$ has no direct contribution from 
the Wess--Zumino term. Using formulae (3.5a) and (B2) of ref 
\cite{Me89b} one verifies that
$\langle r^2\rangle_{I=1} f_\pi^2/g_A$ does not change when 
(\ref{vmscal}) is imposed. This behavior is also obtained in
the pure pseudoscalar Skyrme model.

For completeness we should note that a similar behavior, {\it i.e.}
the non--existence of a universal scaling law is also observed 
for the isoscalar magnetic moment. Hence the 
above discussion applies to the whole isoscalar current.
 
These studies raise the question whether it is possible to construct
a vector meson model without the Wess--Zumino term. Noting that this 
term is crucial not only from a conceptual point of view (chiral 
anomaly) but also for the proper normalization of the nucleon charge 
this question can immediately be answered in the negative.
It turns out that the normalization of the isoscalar charge does not
completely fix the scaling law for the associated radius because the
spatial integral over those parts of the isoscalar density which do
not arise from the Wess--Zumino term vanishes identically \cite{Me89b}.

\bigskip
\leftline{\large\it 4. Local Scaling Laws}
\medskip

The scaling laws (\ref{vmscal}) for the mesonic parameters, which 
were imposed to obtain a universal mass functional, are clearly  
very restrictive and one wonders whether or not other relations can 
be found. While the scaling behavior of the parameters entering 
${\cal L}_{\rm nan}$ acquires some justification from QCD sum rules
there is no a priori analysis which prevents one from choosing scaling
laws different from (\ref{vmscal}) for the parameters describing 
$\Gamma_{\rm an}$. In turn a different choice might yield scaling 
laws for the soliton properties which significantly deviate from 
those obtained in the simple Skyrme model. In contrast to the general 
relations 
given in eq (\ref{vmscal}) such alternative scaling laws will depend 
on the particular values the mesonic parameters take on at zero scaling. We therefore 
refer to these relations as {\it local scaling} laws. As this subject is 
potentially vast we will consider only one example:
\be
{\tilde h}^{\ast}={\tilde h}(1-cx)
\quad {\rm and}\quad
g_{VV\phi}^{\ast}=g_{VV\phi}(1-cx)^2\ ,
\label{lscal}
\ee
while $f_\pi^{\ast}$ and $g^{\ast}$ are taken as given in eq 
(\ref{fgscal}). The unscaled parameters are given in eq (\ref{vmpara}).
The associated Brown--Rho scaling as well as a modified 
Brown--Rho scaling, ${\tilde A}^{\ast}=M^{\ast}/f_\pi^{\ast}g_A^{\ast}$, 
are shown in figure 2.
\begin{figure}[ht]
\centerline{
\epsfig{figure=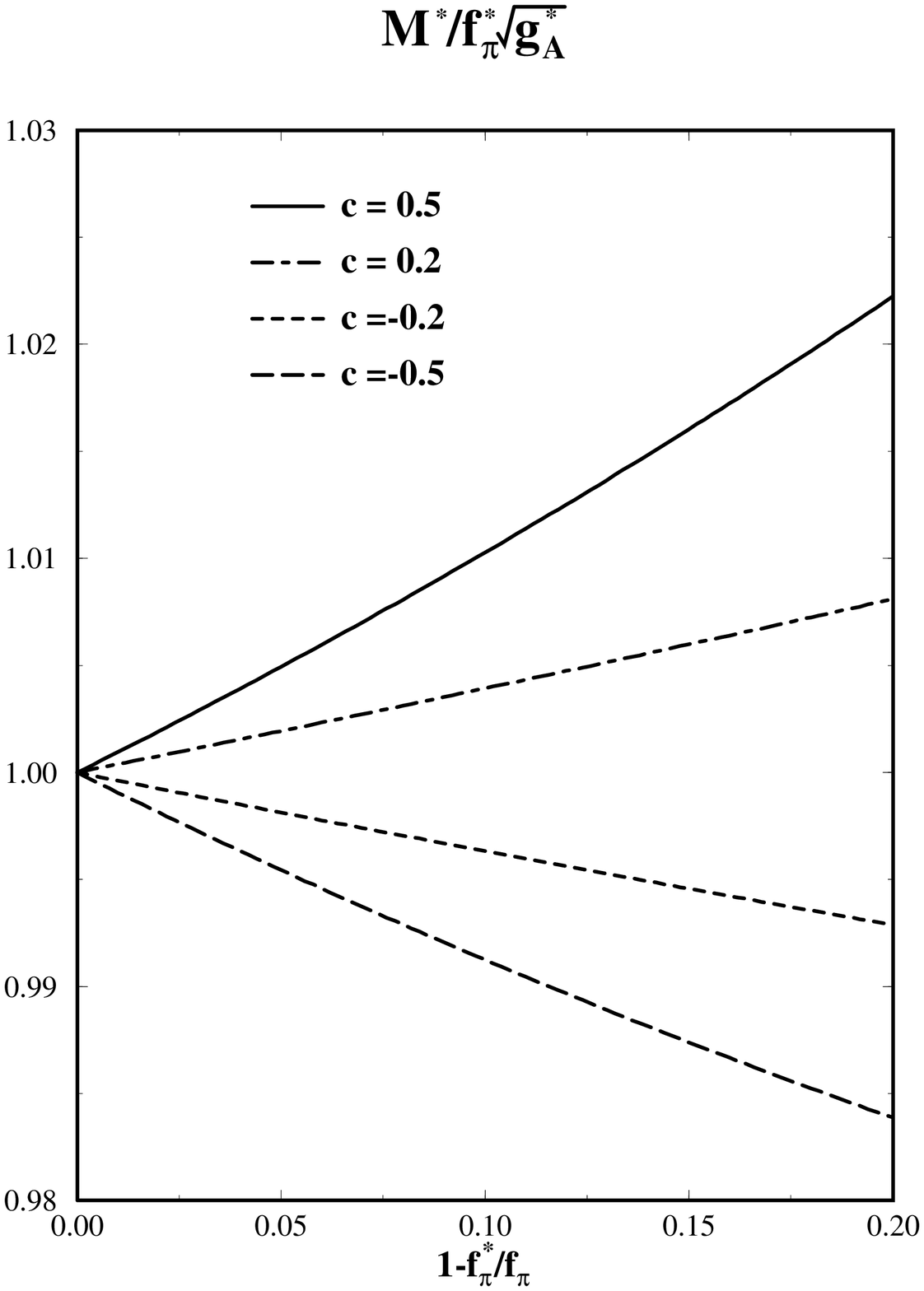,height=8.0cm,width=8.0cm}
\epsfig{figure=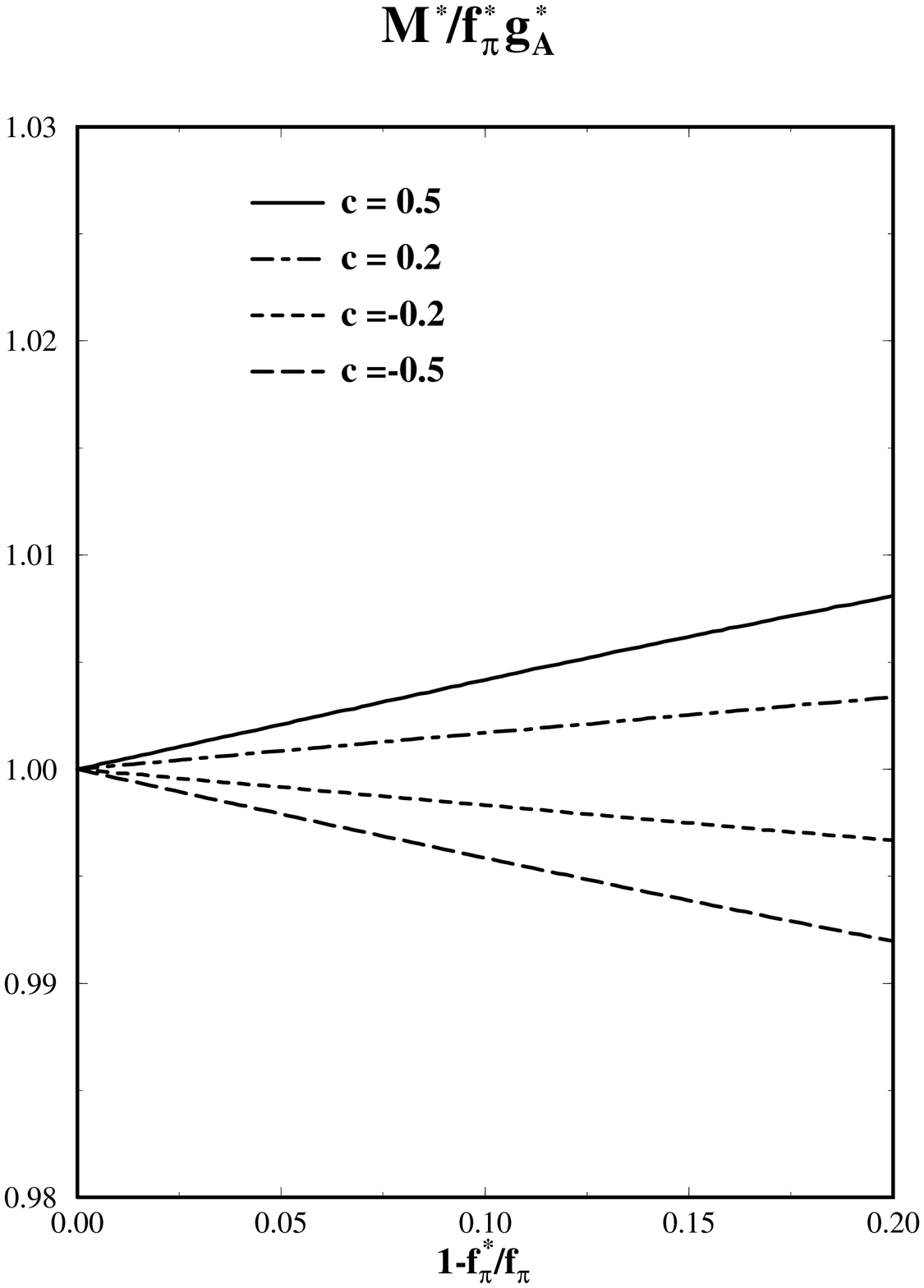,height=8.0cm,width=8.0cm}}
\vskip0.8cm
\caption{Comparison of the two local scaling laws in the 
vector meson model.}
\end{figure}
Although the $x$--dependences of both quantities are very moderate,
${\tilde A}^{\ast}$ apparently varies less than $A^{\ast}$, {\it cf.} 
eq (\ref{defAB}). Certainly a fine--tuning of either the scaling rules 
(\ref{lscal}) or the modified Brown--Rho law could be imposed to 
yield ${\tilde A}^{\ast}$ invariant. This simple example of a local 
scaling law indicates that other scaling behaviors for the baryon 
properties could be justified by a different choice for the variation 
of the parameters entering $\Gamma_{\rm an}$. Hence the confirmation of 
specific scaling laws for static nucleon properties from a realistic 
soliton model seems to be difficult as long as the {\it a priori}
scaling  behavior of the anomalous terms is as poorly known as at 
present. When axial vector mesons are included \cite{Zh94} the 
situation is most likely even more arbitrary.

\newpage

\leftline{\large\it 5. Conclusions}
\medskip

In the framework of chiral soliton models we have studied 
the dependence of static nucleon properties on rescaling of 
the parameters describing the effective meson theory. The first 
step in deriving scaling relations between nucleon properties
and the mesonic parameters consists of constructing a universal 
energy functional for the soliton. In the case of the simple Skyrme 
model this immediately leads to parameter dependences of the 
axial charge and the isosinglet radius which obey the Brown--Rho 
scaling laws. When extending the Skyrme model by 
adding the pion mass term and/or higher order stabilizing terms, 
the scaling laws no longer follow automatically. However, it 
is possible to impose conditions on the additional parameters such 
that the scaling laws are recovered. In the pseudoscalar 
Skyrme model the isoscalar form factor completely arises from 
the Wess--Zumino term, which has no effect on the classical mass 
functional. Therefore the conditions on the mesonic parameters, 
which are derived from the mass functional, have no direct 
consequences for the isoscalar form factor. The situation changes 
drastically when explicit vector meson are contained in the soliton 
model. Then the isoscalar form 
factor receives contributions not only from the anomalous Wess--Zumino
term but also from the direct coupling of the photon to the $\omega$ 
meson and its source terms which are contained in the chirally 
symmetric $\epsilon$--terms. Since these quantities also appear in the 
mass functional there is no unique scaling law for the isoscalar form 
factor. Numerical calculations indicate that the vector terms actually 
dominate the isoscalar form factor, making necessary a modification 
of the Brown--Rho scaling law for the isoscalar radius. In the context 
of the vector meson model we have also seen that a universal mass 
functional requires the KSRF relation to be scale invariant rather 
than just the dimensionless ratio of the vector meson mass over the 
pion decay constant. In addition to this ratio the KSRF relation 
involves the vector meson coupling constant. This coupling constant
has to be taken scale dependent in order to allow the 
axial charge to vary.

These studies were motivated by the derivation of the Brown--Rho 
scaling for the description of dense matter from the simple Skyrme 
model. In general this approach is based on the assumption that the 
density dependence of nucleon properties can be obtained by a 
suitable scaling of the parameters contained in the mesonic 
Lagrangian from which the soliton is constructed. Unfortunately 
only little is known about the density dependence of the 
coefficients of the $\epsilon$ terms. As these play an important 
role for stabilizing the soliton it is not surprising that by 
choosing an arbitrary behavior of these parameters almost any scaling 
law for the nucleon mass can be obtained. We have illustrated this 
for the case that $M/f_\pi g_A$ can be made less sensitive to 
parameter variations than $M/f_\pi \sqrt{g_A}$. When applying these 
results to investigate the density dependence of nucleon properties 
one should also bear in mind that at finite density additional terms 
may be needed in the effective meson theory as, for example, suggested by 
heavy baryon chiral perturbation theory \cite{Be95}.

\vskip1cm
\leftline{\large\it Acknowledgements}
\medskip
One of us (HW) gratefully acknowledges the warm hospitality 
extended to him during a visit at Syracuse University.

This work has been supported in part by the US DOE under contract 
DE-FG-02-85ER 40231 and by the DFG under contracts We 1254/2--2 and 
Re 856/2--2.

\end{document}